# Scaling and Statistics of Bottom-Up Synthesized Armchair Graphene Nanoribbon Transistors

Yuxuan Lin[1,*], Zafer Mutlu[1], Gabriela Borin Barin[2], Jenny Hong[1], Juan Pablo Llinas[1], Akimitsu Narita[3], Hanuman Singh[1], Klaus Müllen[3], Pascal Ruffieux[2], Roman Fasel[2,4], Jeffrey Bokor[1,5,*]

[1] Department of Electrical Engineering and Computer Sciences, University of California, Berkeley, CA, USA

[2] Empa, Swiss Federal Laboratories for Materials Science and Technology, Dübendorf, Switzerland

[3] Max Planck Institute for Polymer Research, Mainz, Germany

[4] Department of Chemistry and Biochemistry, University of Bern, Bern, Switzerland

[5] Materials Sciences Division, Lawrence Berkeley National Laboratory, Berkeley, CA, USA

[*] Correspondence to: yxlin@berkeley.edu, jbokor@berkeley.edu

ABSTRACT

Bottom-up assembled nanomaterials and nanostructures allow for the studies of rich and unprecedented quantum-related and mesoscopic transport phenomena. However, it can be difficult to quantify the correlations between the geometrical or structural parameters obtained from advanced microscopy and measured electrical characteristics when they are made into macroscopic devices. Here, we propose a strategy to connect the nanomaterial morphologies and the device performance through a Monte Carlo device model and apply it to understand the scaling trends of bottom-up synthesized armchair graphene nanoribbon (GNR) transistors. A new nanofabrication process is developed for GNR transistors with channel length down to 7 nm. The impacts of the GNR spatial distributions and the device geometries on the device performance are investigated systematically through comparison of experimental data with the model. Through this study, challenges and opportunities of transistor technologies based on bottom-up synthesized GNRs are pinpointed, paving the way to the further improvement of the GNR device performance for future transistor technology nodes.

INTRODUCTION

Emerging low-dimensional semiconductors have been envisioned as promising candidates for the channel materials of ultimately scaled post-silicon transistor technologies[1-3]. Among them, graphene nanoribbons (GNRs) and their heterostructures produced by on-surface bottom-up synthesis approaches[4-6] are of particular interest, because of their capability of defining the electronic band structures at the limit of the atomic scale, giving rise to intriguing electronic properties, including width-tunable band gaps[7], topologically engineered edge states[8, 9], quantum spin chains[10-12], and steep switching induced by quantum dot and superlattice states[13-16]. Field effect transistors (FETs) made with bottom-up synthesized armchair GNRs have been demonstrated previously[17, 18]. The widths and edges are perfectly defined with atomic resolution and near-unity yield, as indicated in scanning tunneling microscopic (STM) studies[4, 18]. Transport measurements and theoretical analysis have shown that up to 20 µA per GNRs of on-state current with exceptional gate-control capability can be achieved for short-channel FETs representing sub-10 nm technology nodes[19]. However, a number of challenges still remain[20], such as large Schottky barriers at the metal-GNR contacts[18], GNR-GNR bundling[21, 22], short GNR lengths[18, 23], and chemical instability[24, 25]. In addition, although the atomic structures of GNRs are nearly perfectly defined, the lengths, orientations, and locations of GNRs are randomly distributed with the current synthesis approach (see **Figure 1a**). As a result, it has been difficult to correlate the overall device performance with the microscopic features obtained from advanced materials characterization approaches. In this study, we design a methodology that builds connections between the transport measurement and the advanced microscopy results of GNR samples through a Monte Carlo device model. With this approach, we are able to understand the impacts of different geometrical, physical, and process-related factors on the device performance in a quantitative manner. In particular, the dimensions of FETs (channel length and channel width) are systematically varied for different types of 9-atom-wide armchair GNR (9-AGNR) samples. A new nanofabrication process is developed for GNR FETs with channel lengths down to 7 nm. Through a combination of statistical transport measurements, STM and Monte Carlo device simulations on these samples, we find clear correlations between microscopic variables of the GNR samples, such as GNR lengths, densities and spatial distributions, and macroscopic measures of device characteristics, such as the device yield, the on-state current ($I_{ON}$), the on/off current ratio ($I_{ON}/I_{OFF}$), and the subthreshold swing (SS). By comparing the experimental and the simulation results, we find that the GNR-GNR bundling and the GNR-length-limited short contact length are the major limiting factors of the device performance. This approach helps identify the current bottlenecks and future research directions for further improving the performance of GNR transistors and sets a solid foundation for quantitative electrical characterization and device physics studies for bottom-up synthesized nanoscale materials.

RESULTS AND DISCUSSION

Figure 1(a) shows a representative STM image of the as-synthesized 9-AGNRs on a Au(111)/mica substrate. The 9-AGNRs are straight and densely packed on the surface, indicating high quality of the synthesized 9-AGNRs. The 9-AGNRs are then wet-transferred[22] onto a pre-patterned local bottom-gate substrate with 5.5-nm-thick $HfO_2$ gate dielectric for further device fabrication and transport characterization (see Methods for details about synthesis, transfer, and device fabrication). To verify the fidelity of the transfer process, Raman spectroscopy measurements are performed on the samples before and after the transfer (blue and red in Figure 1(b), respectively), with an excitation laser wavelength of 785 nm. The sharp and strong peak at 314 cm$^{-1}$ corresponding to the radial breathing-like mode (RBLM) is well-preserved after transfer, suggesting that the atomic structures of the 9-AGNR samples[26, 27] remain intact. After transfer, the source-drain metal contacts (Pd) are patterned by a process that involves e-beam lithographically defined local shadow masks and two steps of tilted e-beam evaporation (see Methods). With this process, it is possible to resolve channel lengths down to 7 nm. The device schematic and the top-view scanning-electron microscopic (SEM) images of the as-fabricated devices are shown in Figure 1c, d and Figure S1.

Figure 1e, and 1f represent typical transport characteristics of the as-fabricated 9-AGNR FET with a channel length (*L*) of 20 nm and a channel width (*W*) of 500 nm. According to the transfer characteristic

(drain current, $I_D$, versus gate voltage, $V_{GS}$) as shown in Figure 1e, p-type FET behavior is observed, with $I_{ON}$ of 3.5 µA at a drain voltage ($V_{DS}$) of 1 V, and $I_{ON}/I_{OFF}$ on the order of $10^3$, which are among the best reported values for bottom-up synthesized GNR transistors[15, 17, 18, 28-30]. Given that the band gap of 9-AGNR has been theoretically predicted to be ~2.2 eV[7], it is in general challenging to form ohmic contact to semiconductors with such a large band gap. The super-linearity of the output characteristics ($I_D$ versus $V_{DS}$) as shown in Figure 1f indicate the presence of a finite Schottky barrier at the metal-GNR interface, however it is considered to be more linear than previously reported results[15, 17, 18, 28-30], indicating an improved metal contact. This improvement is likely due to the larger average GNR lengths, which will be discussed later.

To investigate the impact of the length scales of GNRs on the device performance, 9-AGNR samples with various average lengths and densities are synthesized with different growth conditions, and the statistics about the length and spatial distributions are characterized with STM and captured by a Monte Carlo model. The right panels of **Figure 2a-c** display typical STM images of 9-AGNRs on Au(111)/mica substrates synthesized under low-coverage (LC, 0.3 monolayer), medium-coverage (MC, 0.5 monolayer), and high-coverage (HC, 1 monolayer) conditions. The GNR length ($l_{GNR}$) distributions for these three types of samples are extracted from multiple STM images as shown in Figure 2d-f. The means and standard deviations are 17.29 nm and 11.77 nm for LC samples, 19.70 nm and 13.19 nm for MC samples, and 47.87 and 29.20 nm for HC samples. The densities of GNRs (defined as numbers of GNRs per unit area) are 0.0147 nm$^{-2}$, 0.0273 nm$^{-2}$, and 0.0188 nm$^{-2}$, respectively. It is observed that the LC and MC samples have similar GNR length distributions, while the MC and the HC samples are more densely packed than the LC sample. The asymmetric length distributions can be fitted well with the gamma distribution as shown in red solid curves in Figure 2d-f. The spatial distributions can be further modeled by a Monte Carlo simulation with the extracted length distributions, the GNR densities and the minimum GNR-to-GNR separations as the input parameters (see Methods for details). As shown in the left panels of Figure 2a-c, the simulated spatial distributions can capture the main features observed in experimental results (STM images, right panels of Figure 2a-c) reasonably well, especially in terms of the short-range alignment and the long-range randomness.

Raman spectroscopy measurements are taken on these three types of samples before and after transfer and device fabrication, which provides complementary valuable information about the GNRs at macroscopic scale. Figure S2a-c displays the typical Raman spectra of the HC, MC, and LC GNRs as-grown on Au(111)/mica and the GNR devices, respectively. The presence of the well-defined characteristic RBLM, C-H mode, D, and G modes for all samples confirms the successful growth and transfer of the 9-AGNRs on Au(111)/mica[22]. When compared, the HC GNR sample has the sharpest and highest intensity RBLM peak, which indicates its high quality[22]. This result is well-consistent with the results of a recent study[31] reporting that when the molecules are more densely packed on the gold surface, the GNR quality in terms of the length and alignment is improved significantly due to the increased polymerization reaction and the reduced premature cyclodehydrogenation. The HC and MC samples are also more stable under potential processing environments[32] since the neighboring GNRs protect each other from the environmental contaminants while at the low coverage the GNR edges are more exposed to them. Moreover, while the intensities of the RBLM, G, D, and CH peaks in the Raman spectra of the HC and MC devices are almost similar, the peak intensities of the LC device are lower than those of the two devices. This result is expected since the Raman intensities are proportional to the GNR density, and it further confirms that the GNRs are seamlessly transferred.

With the help of the statistics of GNRs in the microscopic scale as discussed above, we can now better understand the statistics and scaling effects of the transport characteristics. Figure 2g-i are representative transfer characteristics of FETs made with the three types of 9-AGNR samples, with the channel lengths ranging from 7 nm to 90 nm. To confirm that the measured $I_D$ are the current flowing through the GNR channels, instead of the gate leakage current ($I_G$), or the short-circuit current across the source-drain electrodes due to fabrication imperfections, we measured the $I_G$ as well as $I_D$ on a device with similar dimensions, but without GNRs in the channel, which are shown in Figure S3. Both currents are several orders of magnitude smaller than the $I_D$ measured on the 9-AGNR FETs, which rules out such possibilities. Clear correlations are observed in Figure 2g-i among the device performance, the FET channel length, and the types of 9-AGNR samples. To better capture the channel length scaling effect, we fabricate and measure tens of 9-AGNR FETs for each dimension and each GNR coverage and summarize the statistics of the key device performance metrics.

First, dramatic increases of both $I_{ON}$ and the device yield are observed as $L$ is downscaled. The scattered points in **Figure 3a** and b are $I_{ON}$ versus $L$ for devices made with different types of 9-AGNR samples (plotted in different colors) and in different $W$ (100 nm and 500 nm, respectively). In addition, the device yield is also improved as $L$ is reduced (Figure 3e and f). Here, the device yield is defined as the percentage of devices, at each dimension, with their measured $I_D$ at least 10 times larger than their measured gate leakage current ($I_G$). This yield improvement is more substantial for the LC and MC samples. Both the trends for $I_{ON}$ and for device yields can be explained by the larger number of connected GNRs and the longer average contact length (defined as the overlap between the GNR and the source/drain metal contact; labeled at $L_S$ and $L_D$ in Figure S4) of GNR-metal interfaces as the channel lengths decrease, given certain length and spatial distributions of the 9-AGNR samples. Here the contact length is still smaller or comparable to the characteristic length of the current "crowding" at the contact edge (called transfer length), which is a major limiting factor for the contact resistance of 9-AGNR FETs. This effect has also been observed in carbon nanotubes [33].

To quantify the impacts of the device dimension and the GNR distribution on the device performance, a simplified device model is developed based on the Monte Carlo simulation results as discussed earlier. Figure S4 illustrates a simulated GNR spatial distribution, with the box in purple dashed line indicating the active device area. Then it is straightforward to find how many GNRs are connected to both the source and the drain electrodes for each simulated device. These are the GNRs that contribute to the channel conductance. This simulation process is repeated for multiple times at each device dimension, and the resulting median numbers of connected GNRs ($N$) as a function of $L$ for different types of 9-AGNR samples are plotted in Figure 3c, d and Figure S5a-c. Meanwhile, the simulated device yield can be obtained if devices with $N > 0$ are considered as working devices. The simulated device yields are shown in dashed lines in Figure 3e and f, and additional results are shown in Figure S5d-f. An interesting observation is that the simulated and the experimental results match with each other to different extent for different samples and different device dimensions. In particular, they agree with each other very well for the LC 9-AGNR sample among various $L$ and $W$, whereas the measured yields are higher than the simulated ones for the MC and HC samples when $L$ is larger and $W$ is smaller ($W = 100$ nm in Figure 3e). Given that the MC and HC samples have higher densities than the LC sample, we infer that the larger difference between the simulated and the experimentally extracted yields for MC and HC samples may originate from the additional inter-ribbon conductance due to GNR relocation. An underlying hypothesis for the Monte Carlo simulation is that the spatial distributions of GNRs are preserved after the device fabrication process, and as a result, no inter-ribbon electrical conductance is considered. This hypothesis may not be accurate, as relocation of GNRs may take place especially during the wet-transfer process, and the neighboring GNRs are likely to contact, or bundle with one another, leading to the inter-ribbon conductance[22]. Assuming similar degrees of GNR relocation for the three types of 9-AGNR samples, the higher density, or lower GNR-GNR spacing in MC and HS samples would give rise to a higher probability of physical contact of neighboring GNRs and more contributions from the inter-ribbon conductance. Therefore, the higher yields measured on devices with larger $L$ and smaller $W$ for these two types of samples can be attributed to the longer-range and winding conduction paths of interconnected GNRs within and outside the channel regions, respectively.

The total on-current $I_{ON}$ for each device can be computed by the summation of the current passing through each GNR that is connected to the source and drain electrodes. To estimate the single-GNR current, a simplified model is used based on the Landauer formula[34] with the considerations of the transmission probability through the source/drain contact barriers and the effect of short contact length (see Methods for details). Figure 3a and b display the simulated and experimental average $I_{ON}$ for the three types of 9-AGNR samples when $W = 100$ nm and 500 nm, respectively, and they are in reasonable agreement with each other. Figure S5g-i show additional simulated $I_{ON}$ with various device dimensions. We thus conclude that the negative correlation between $I_{ON}$ and $L$ are mainly because of the change in the number of connected GNRs in the channel, as well as the change in the contact length. In addition, the slightly faster $I_{ON}$ drop with $L$ in the shorter $L$ regime ($L < 20$ nm) observed in experiments may come from defect scattering in the channel[35], and the more gradual tail in the longer $L$ regime ($L > 50$ nm for the HC sample) may be another indication of the inter-ribbon conductance as mentioned earlier. In addition, the much lower $I_{ON}$ for the LC sample can be attributed to both the smaller number of connected GNRs and the lower GNR quality than those of HC and MC samples as substantiated by the Raman spectroscopy results as discussed earlier (Figure S2).

Second, the gate coupling efficiency is also strongly correlated with the channel length, as suggested by the scaling trends for both the $I_{ON}/I_{OFF}$ and the SS as shown in **Figure 4a** and b, respectively. Here, SS is defined as SS = $|d \lg(I_D)/dV_{GS}|^{-1}$, and the minimum SS among different $V_{GS}$ is selected to represent the performance of each device in Figure 4b. The average $I_{OFF}$ as a function of $L$ are also shown in Figure S6a. At the long-channel limit, the average $I_{ON}/I_{OFF}$ is as high as $10^4$, and the average SS is as low as 100 mV/dec, which indicate good gate coupling efficiencies. However, when $L$ becomes smaller, the average $I_{ON}/I_{OFF}$ decreases exponentially and the average SS increases exponentially, corresponding to a fast degradation of the gate control. The transition between the flat and the fast-changing regimes happens at the critical channel length $L_{cr} \approx 30$ nm. However, for an ideal single-GNR channel on a 5.5-nm-HfO$_2$ gate dielectric, the scaling length $\lambda$ is estimated to be only ~2 nm[36], which is much lower than $L_{cr}$. Such a discrepancy could be attributed to two factors: (1) GNR-GNR screening, in which the neighboring GNRs in a densely packed sample screen the fringe capacitance of each GNR channel, leading to a weakened gate coupling efficiency; (2) GNR relocation and bundling, in which there might be a chance that neighboring GNRs are on top of each other, resulting in a higher effective body thickness. In future work, aligned GNRs with low density[37] and dry transfer techniques can be adapted to address these issues. Furthermore, the bottom-gate device geometry can be replaced by a dual-gate or a gate-all-around geometry to further improve the gate coupling efficiency and to enable the further downscaling of GNR transistors[36, 38, 39].

Third, the source/drain metal contact to the GNR channel is improved for the longer GNR samples (HC 9-AGNR). As shown in Figure 4c, the $I_D$-$V_{DS}$ curve for the MC sample is very nonlinear, indicating a large Schottky barrier tunneling resistance at the metal-semiconductor contact, whereas the $I_D$-$V_{DS}$ curve for the HC sample with the same device dimension is almost linear, indicating a better metal contact. To quantify this effect, we define the nonlinearity NL as NL = $I_D(V_{DS} = -1 \text{ V})/[10 I_D(V_{DS} = -0.1 \text{ V})]$: larger NL (greater than 1) means stronger superlinear $I_D$-$V_{DS}$ relation; and NL = 1 corresponds to linear $I_D$-$V_{DS}$ relation. The average NL for the three types of 9-AGNR samples for various $L$ and $W$ are plotted in Figure 4d. A systematically lower NL is observed on the HC 9-AGNR sample than on the LC and MC samples, for the same $L$. This trend is further substantiated by the systematically lower drain-induced barrier lowering (DIBL) for the HC sample as shown in Figure S6b. We thus conclude that such a nonlinearity improvement for the HC sample is associated with the larger contact length, or the larger overlap between the GNR channels and the source/drain electrodes.

## 3. Conclusion

In summary, through varying the device geometry and extracting the statistical trends of the device parameters, the correlations between the GNR microscopic morphologies and the device performance of the GNR FETs are obtained. With a combined theoretical and experimental approach, the effects of the GNR spatial distributions, the GNR-GNR relocation and bundling during the device fabrication processes, and the overlaps between the metal contacts and the GNR channel are quantitatively investigated. Based on our analysis, we identify the factors limiting GNR devices fabricated using the present technologies, and anticipate that further improvement of the device performance be achieved for aligned GNR samples with longer GNR lengths and controlled GNR densities.

EXPERIMENTAL SECTION

**Materials Synthesis and Characterization**

*9-AGNR Synthesis.* 9-AGNR samples were produced in a stand-alone ultrahigh-vacuum (UHV) system dubbed the "GNR reactor" - a fully automated system that allows reproducible and high-throughput fabrication of GNRs[22]. A Au(111)/mica substrate (Phasis, Switzerland) was cleaned by two cycles of Ar+ sputtering (1 keV, for 10 minutes) and annealing (470 ºC, for 10 minutes). 9-AGNRs were synthesized by thermal sublimation of of 3′,6′-diiodo-1,1′:2′,1″-terphenyl (DITP)[23] followed by two sequential annealing steps at 200 ºC and 400 ºC to activate polymerization and cyclodehydrogenation (CDH) reactions, respectively. The three molecular coverages investigated in this work were achieved by controlling both the precursor sublimation rate with a quartz microbalance and the deposition time.

*Raman Spectroscopy Measurement.* Raman characterization of the GNRs was performed using a Horiba Jobin Yvon LabRAM ARAMIS Raman microscope using a 785 nm laser with <10 mW power and a 100× objective lens, resulting in a laser spot size of <1 μm. No thermal effects were observed under these measurement conditions, and at least three spectra from different points were collected for each sample to verify the consistency. The measurements were taken at ambient conditions.

*Scanning Tunneling Microscopy (STM).* Topographic STM images of as-grown 9-AGNRs on Au(111)/mica samples were taken with a Scienta Omicron VT-STM operated at room temperature. Constant-current STM images were recorded. The sample bias and the set point current are indicated in the associated figure captions.

**Device Fabrication and Measurement**

*Preparation of Local Bottom Gate Substrate.* The starting substrate is 100 nm $SiO_2$/Si. The local back gates are ~8 nm W deposited through sputtering, and lithographically patterned and wet etched by $H_2O_2$. The ~5.5 nm $HfO_2$ was grown in an atomic layer deposition (ALD) system (Oxford, FlexAl Plasma ALD) at 135 °C. Alignment markers and large pads for electrical probing were patterned using standard photolithography and lift-off of ~3 nm Cr and ~25 nm Pt. The wafer was then diced, and individual chips were used for further device processing.

*GNR Transfer.* First, the GNRs grown on the Au(111)/mica substrates were floated in 38% HCl in water, which caused the substrates to delaminate with the GNRs/Au film floating on the surface of the acid. Next, the floating GNRs/Au film was picked up with a local bottom gate substrate, with the GNRs facing the substrate. To increase adhesion between the gold films and substrates, a drop of isopropanol was placed onto the gold thin films (dried at ambient conditions for 5 min) followed by a hot plate baking at 100 °C for 10 min. A gold etchant solution (potassium iodine, no dilution) was used to etch the remaining gold layer. After ~5 min of etching, the samples were cleaned by soaking them in deionized water for 10 min followed by rinsing with isopropanol and dried under a stream of nitrogen.

*Patterning of Source/Drain Electrode.* After GNR transfer, the samples were spin-coated with double-layer e-beam resist (methyl methacrylate (MMA)/ poly(methyl methacrylate) (PMMA)) and patterned by e-beam lithography (Crestec CABL-UH 130 kV). After development, the MMA layers were undercut, leaving suspended PMMA stripes with 100-200 nm widths as local shadow masks. Then two consecutive steps of electron-beam evaporation of 10 nm Pd with the samples tilted at the opposite angles were performed, followed by a lift-off process, leading to 10-100 nm Pd nanogaps as the channels of the FETs.

*Electrical Characterization.* The electrical characterization of the GNR FETs was performed in a Lakeshore TTPX cryogenic probe station with the vacuum level of <$10^{-5}$ torr, using an Agilent B1500 semiconductor parameter analyzer. All the measurements were performed at room temperature.

**Device Modeling**

*Monte Carlo Model for GNR spatial distributions.* The length of GNRs can be represented by the gamma distribution: $l_{GNR}$ ~ gamma($\alpha, \beta$), with the probability distribution function (PDF) expressed as $\text{PDF}(l_{GNR}) = \beta^{-\alpha} \Gamma^{-1}(\alpha) l_{GNR}^{\alpha-1} \exp(-l_{GNR}/\beta)$, where $\Gamma(\cdot)$ is the gamma function; $\alpha$ and $\beta$ are fitting parameters. To generate the spatial distribution of GNRs, we first determine the total number of GNRs that need to be placed on top of a given area, $N_{tot} = \rho \cdot A$, where $\rho$ is the GNR density extracted from the STM images, and $A$ is the area to be simulated. Then the lengths of the $N_{tot}$ GNRs are generated randomly according to the gamma distribution: $l_{GNR} \sim \text{gamma}(\alpha, \beta)$. Finally, each generated GNR is placed randomly onto the simulation area sequentially based on either of the following two scenarios with certain probabilities: (a) with the probability $p_a$, the current GNR is placed close to the previous one, with the center-to-center distance $\Delta c$ generated randomly from a Gaussian distribution: $\Delta c \sim \text{Gaussian}(0, \sigma_c)$, and the relative angle between the current and the previous GNRs $\Delta\theta$ also generated from a Gaussian distribution: $\Delta\theta \sim \text{Gaussian}(0, \sigma_\theta)$; (b) with the probability $p_b = 1 - p_a$, the current GNR is placed completely randomly (for both center position and the angle) onto the simulation area. An additional constraint needs to be met for both scenarios: the current GNR cannot overlap or be placed closer to any of the already placed GNRs than the minimum GNR-to-GNR separation, $S_{min}$. These two scenarios are chosen based on the following observations: (a) there is local alignment between neighboring GNRs with relatively small rotation angles; (b) at a larger length scale, the locally aligned GNRs form domain structures: within each domain, the GNRs are aligned, whereas in between different domains, the GNR angles are completely random. For LC samples, $\alpha = 2.47$, $\beta = 7.00$, $\rho = 0.0147$ nm$^{-1}$, $\sigma_c = 2$ nm, $\sigma_\theta = 3°$, $p_a = 0.999$, and $S_{min} = 2$ nm; for MC samples, $\alpha = 2.44$, $\beta = 8.05$, $\rho = 0.0273$ nm$^{-1}$, $\sigma_c = 2$ nm, $\sigma_\theta = 3°$, $p_a = 0.999$, and $S_{min} = 2$ nm; for HC samples, $\alpha = 3.50$, $\beta = 15.23$, $\rho = 0.0188$ nm$^{-1}$, $\sigma_c = 10$ nm, $\sigma_\theta = 3°$, $p_a = 0.9997$, and $S_{min} = 2$ nm.

*Estimation of the On-State Current.* The total current ($I_{ON}$) can be computed by a summation of the current passing through each GNR that is connected to both source and drain ($I_n$): $I_{ON} = \sum_n I_n$. From the Monte Carlo simulation, we can find the connected GNR and its dimension parameters (length of GNR, $l_{GNR}$; the rotation angle of GNR with respect to the channel direction, $\theta$, as well as the contact length on the source and drain side, $L_S$, and $L_D$) as indicated in the red line in Figure S4. Also, the effective channel length for this GNR is different from the channel length of the device, $L$, because of the rotation angle $\theta$, which can be expressed as $l = L/\cos(\theta)$. Assuming that the carrier scattering within the GNR channel is negligible, $I_n$ can be estimated according to the Landauer formula[34, 40, 41]:

$$I_n \propto \frac{T_S T_D}{T_S + T_D - T_S T_D} \tag{1}$$

where $T_{S(D)}$ is the transmission probability through the tunneling barrier at the source(drain) contact. Considering the finite contact length $L_{S(D)}$, $T_{S(D)}$ can be estimated by the transmission line model[42]:

$$T_{S(D)} = T_{C0} \tanh\left(\frac{L_{S(D)}}{L_T}\right) \tag{2}$$

where $T_{C0}$ is the transmission probability at the contact when the contact length is infinitely long; and $L_T$ is the transfer length, with an estimated value of 25 nm according to a previous work on carbon nanotubes[43]. Figure S5g-i summarize the calculated $I_{ON}$ for different device dimensions and different types of 9-AGNRs based on the GNR spatial distributions obtained by the Monte Carlo simulation. In Figure 3a and b, the simulated $I_{ON}$-$L$ curve for each type of 9-AGNRs is multiplied by a constant to roughly match the magnitudes of the experimental results.

AUHOR INFORMATION

**Author Contributions**

J.B. supervised the project; Y. L., Z. M. and J. B. conceived the experiments; Y. L., Z. M., J. P. L. and H. S. carried out the device fabrication; Y. L. and Z. M. performed the electrical characterization and data analysis; Y. L. conducted the device modeling; G. B. B. synthesized the materials and performed the STM characterization supervised by P. R. and R. F.; A. N. synthesized the precursors supervised by K. M.; Y. L. conducted the SEM measurements; Z. M. and J. H. performed the Raman spectroscopy measurements. Y. L., Z. M., G. B. B. and J. B. wrote the manuscript with the inputs from all the authors.

**Notes**

The authors declare no competing financial interest.

ACKNOWLEDGEMENT

This work was supported in part by the Office of Naval Research (ONR) MURI Program N00014-16-1-2921, the National Science Foundation (NSF) Center for Energy Efficient Electronics Science (E3S), and the NSF under award DMR-1839098. G.B.B, P.R. and R.F a acknowledge funding by the Swiss National Science Foundation under grant no. 200020_182015, the European Union Horizon 2020 research and innovation program under grant agreement no. 881603 (GrapheneFlagship Core 3), and the Office of Naval Research BRC Program under the grant N00014-18-1-2708. Additional support was provided by the Berkeley Emerging Technology Research (BETR) Center and Taiwan Semiconductor Manufacturing Company (TSMC). Device fabrication was mostly performed at the Marvell Nanofabrication Laboratory at the University of California, Berkeley (UCB). Raman spectroscopy characterization and part of the device fabrication were performed at the Molecular Foundry at Lawrence Berkeley National Laboratory (LBNL), supported by the Office of Science, Office of Basic Energy Sciences, of the U.S. Department of Energy (DOE) under contract no. DE-AC02-05CH11231. We would like to acknowledge fruitful discussions with Prof. P. Kim at Harvard University that motivated this work. We thank X. Hu and Prof. T.-J. K. Liu at UCB for assisting with the transport measurements, as well as D. Dai, C. Su, Prof. A. Zettl at UCB, and S. Aloni at LBNL for assisting with electron microscope characterizations. We also thank E. Chan, S. Dhuey, S. Shelton, T. Mattox, and A. Schwartzberg for laboratory and instrument access in the Molecular Foundry at LBNL.

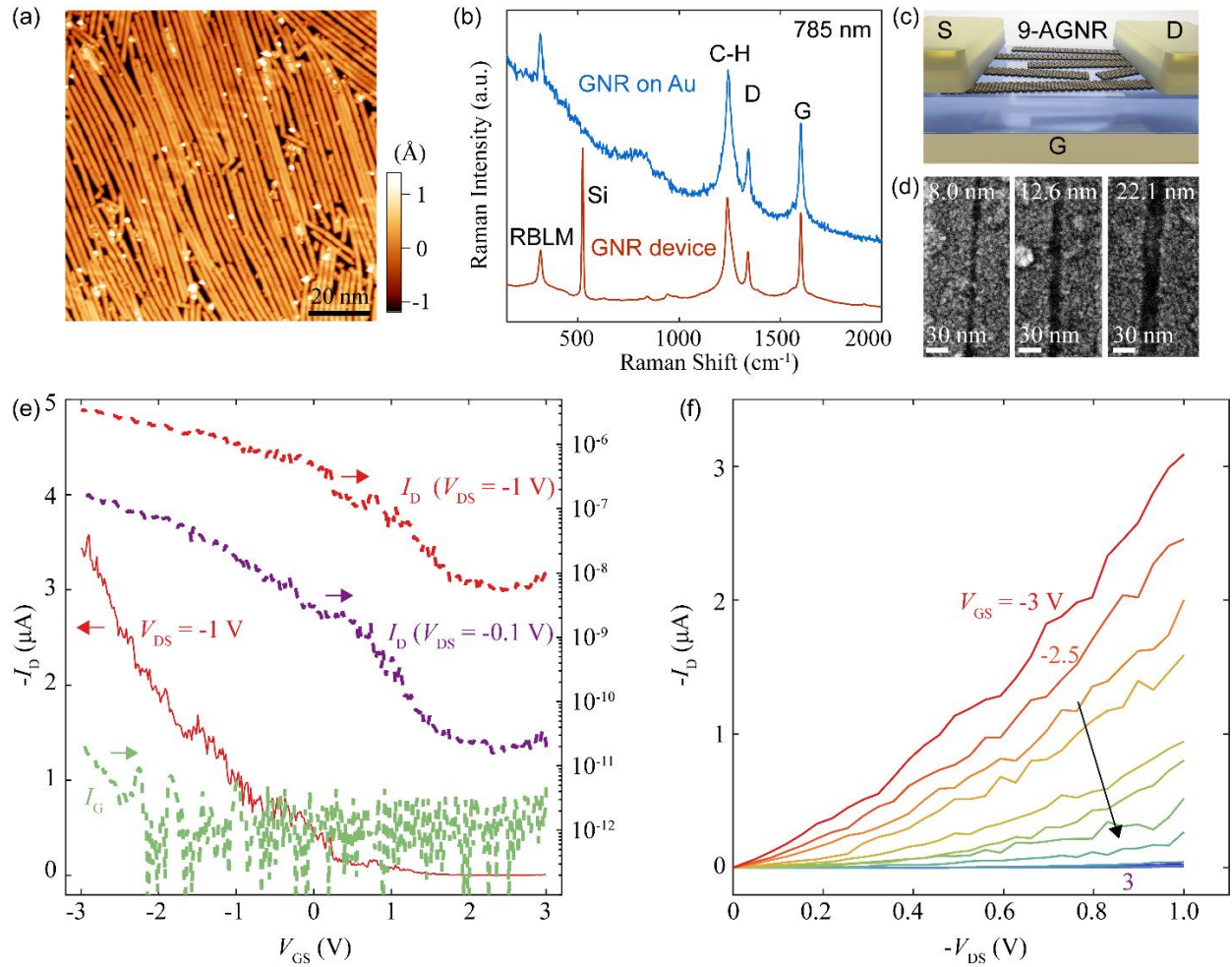

Figure 1 FETs made with bottom-up synthesized 9-atom wide armchair graphene nanoribbons (9-AGNR). (a) STM image of synthesized 9-AGNR (with 1 monolayer coverage) on a Au/mica substrate with sample bias voltage ($V_s$) of -1.5 V and setpoint current ($I_t$) of 5 pA. (b) Raman spectra of 9-AGNR as synthesized on Au (blue) and after transferred onto the HfO$_2$ local bottom gate substrate. The laser excitation wavelength is 785 nm. (c) Schematic of the 9-AGNR FETs with tungsten local back gate, 5 nm HfO$_2$ gate dielectric and Pd source/drain electrodes. (d) SEM images of the as-fabricated devices with different metal gaps (channel lengths) labeled on the images. (e) $I_D$ – $V_{GS}$ characteristic of a 9-AGNR FET in linear scale (left axis) and in logarithmic scale (right axis). The channel length (*L*) and the channel width (*W*) are 20 nm and 500 nm, respectively. (f) $I_D$ – $V_{DS}$ characteristic of the same device with $V_{GS}$ varying from -3 V to 3 V with 0.5 V steps.

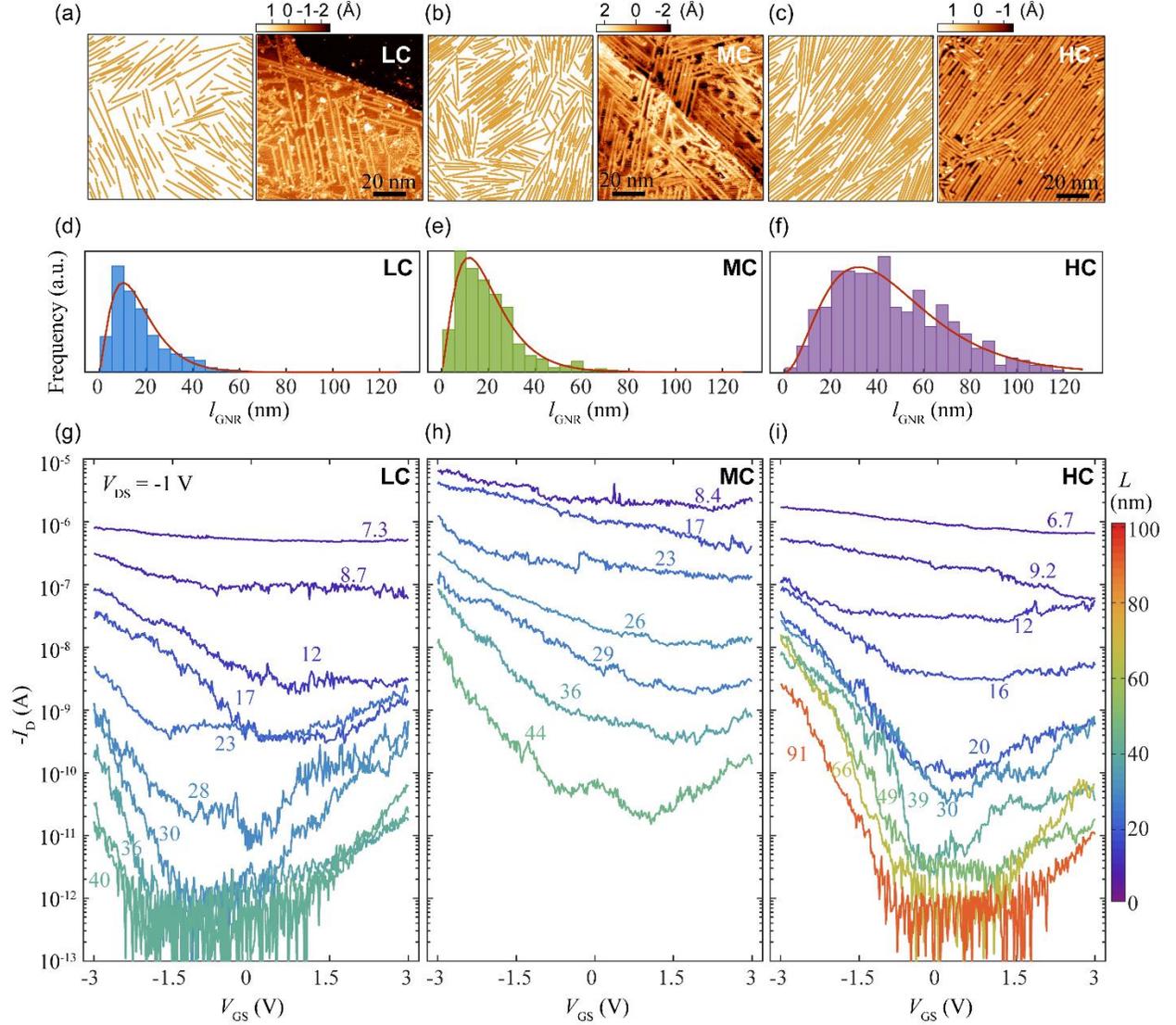

Figure 2. GNR spatial distributions and channel length scaling. (a-c) Representative simulated (left) and experimentally obtained (right) STM images of LC (a), MC (b) and HC (c) 9-AGNRs on Au/mica substrates. $V_s$ = -1.5 V; $I_t$ = 20, 30, and 5 pA, respectively. (d-f) length distributions of LC (d), MC (e) and HC (f) 9-AGNRs extracted from multiple STM images. They are fitted with gamma distributions as shown in red curves. (g-i) Typical $I_D$-$V_{GS}$ characteristics of FETs made with LC (g), MC (h) and HC (i) 9-AGNRs with $W$ = 500 nm and different $L$. The curves are color-coded and labeled with $L$ in nm.

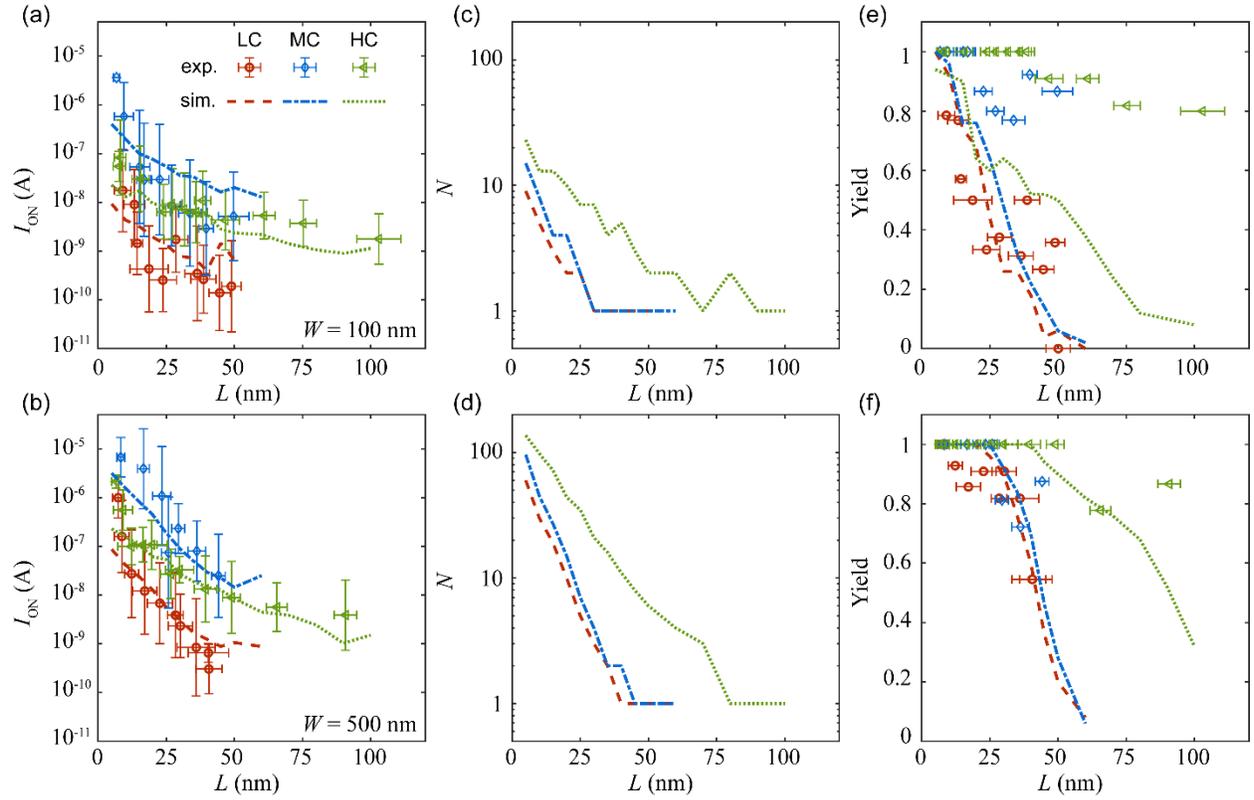

Figure 3. Channel length dependence of $I_{ON}$, $N$, and device yield. (a) $I_{ON}$ versus $L$ for devices with $W = 100$ nm. (b) $I_{ON}$ versus $L$ for devices with $W = 500$ nm. (c) $N$ versus $L$ for devices with $W = 100$ nm. (d) $N$ versus $L$ for devices with $W = 500$ nm. (e) Yield versus $L$ for devices with $W = 100$ nm. (f) Yield versus $L$ for devices with $W = 500$ nm. The scattered points are experimental results for HC, MC and LC GNR samples. Each point is extracted from at least 10 devices with the same dimension. The dashed lines are the corresponding simulation results.

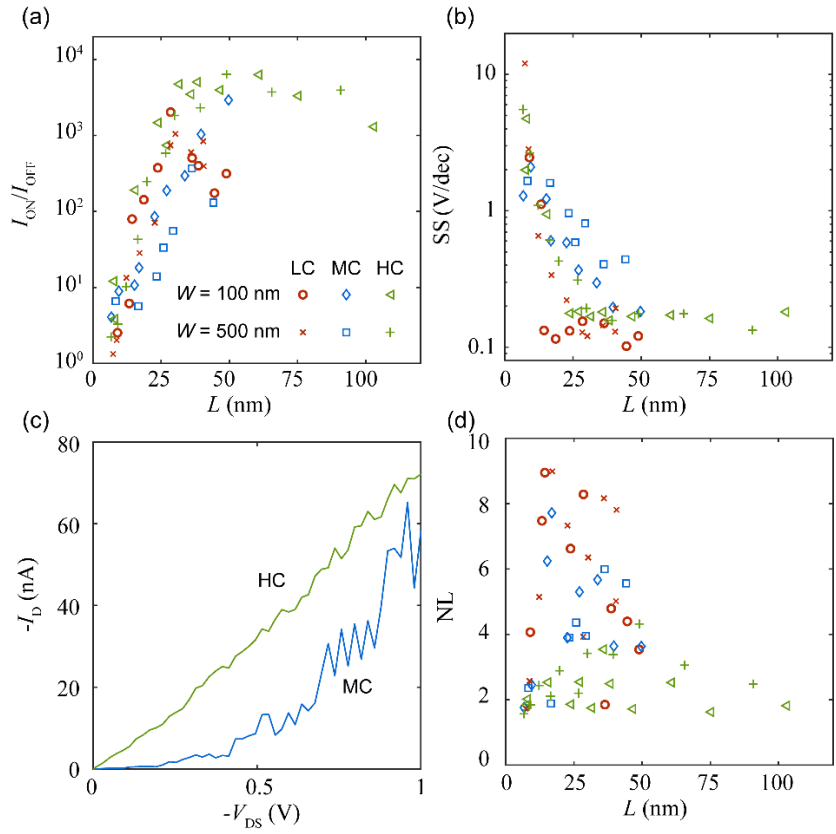

Figure 4. Channel length dependence of $I_{ON}/I_{OFF}$, SS, and NL. (a) Average $I_{ON}/I_{OFF}$ versus $L$ at $V_{DS}$ = -1 V. (b) Average SS versus $L$ at $V_{DS}$ = -0.1 V. (c) $I_D$ – $V_{DS}$ curves of typical MC and HC devices with $L$ = 23 nm, $W$ = 500 nm, at $V_{GS}$ = -3 V. (d) Average NL versus $L$ at $V_{GS}$ = -3 V.

Supporting Information

# Scaling and Statistics of Bottom-Up Synthesized Armchair Graphene Nanoribbon Transistors

Yuxuan Lin[1,*], Zafer Mutlu[1], Gabriela Borin Barin[2], Jenny Hong[1], Juan Pablo Llinas[1], Akimitsu Narita[3], Hanuman Singh[1], Klaus Müllen[3], Pascal Ruffieux[2], Roman Fasel[2,4], Jeffrey Bokor[1,5,*]

[1] Department of Electrical Engineering and Computer Sciences, University of California, Berkeley, CA, USA

[2] Empa, Swiss Federal Laboratories for Materials Science and Technology, Dübendorf, Switzerland

[3] Max Planck Institute for Polymer Research, Mainz, Germany

[4] Department of Chemistry and Biochemistry, University of Bern, Bern, Switzerland

[5] Materials Sciences Division, Lawrence Berkeley National Laboratory, Berkeley, CA, USA

[*] Correspondence to: yxlin@berkeley.edu, jbokor@berkeley.edu

KEYWORDS

Graphene nanoribbon, field-effect transistor, nanoelectronic device, low-dimensional materials

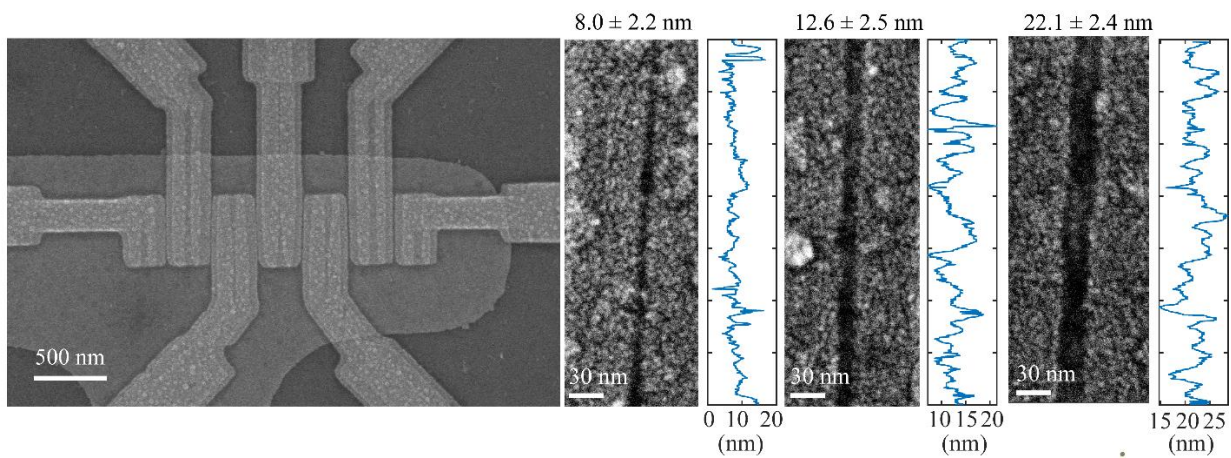

Figure S1. Scanning-electron microscopic (SEM) images and the extracted gap profiles of the metal nanogaps fabricated through the local shadow mask and tilted deposition method.

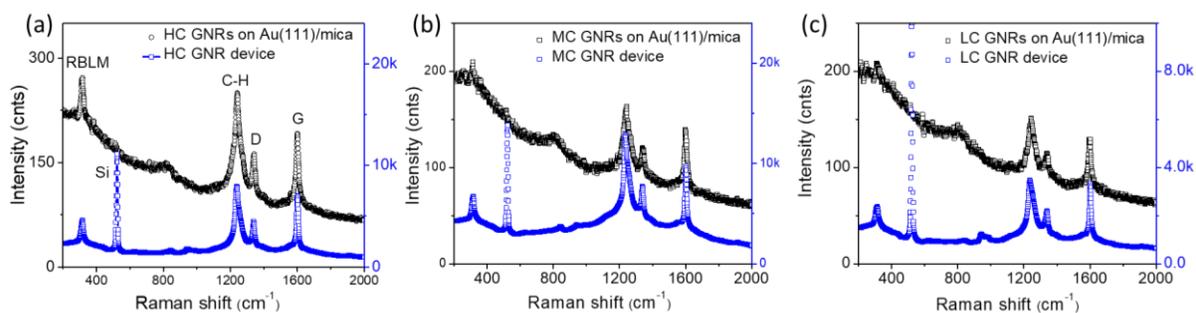

Figure S2. Typical Raman spectra of the HC (a), MC (b), and LC (c) 9-AGNRs as-grown on Au(111)/mica and after transfer and device fabrication.

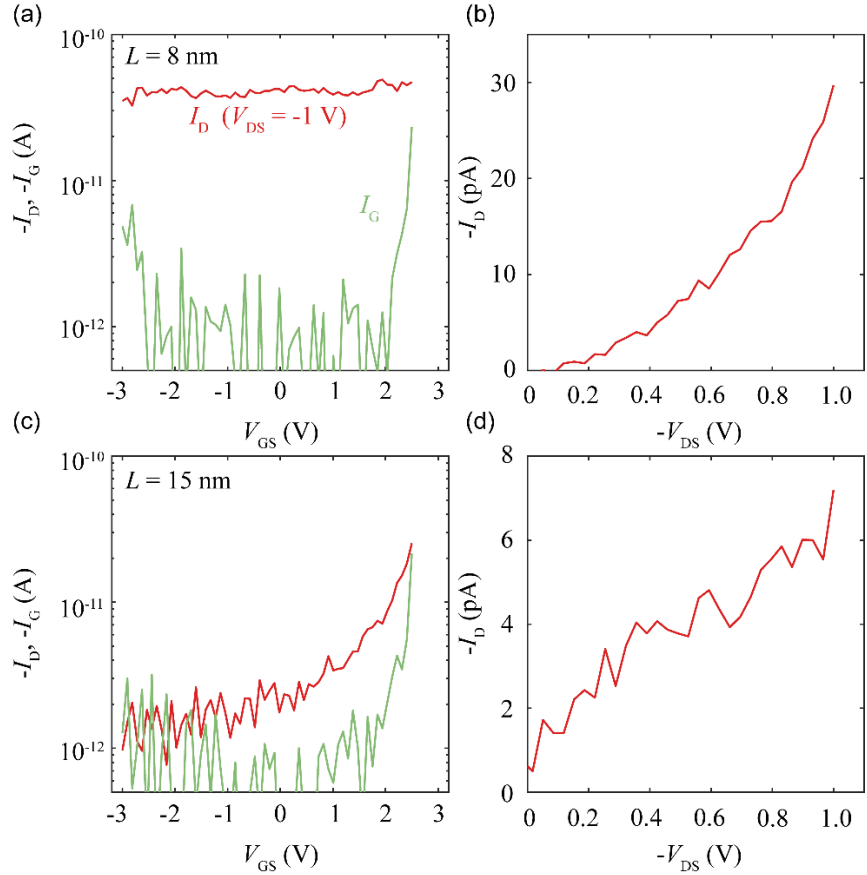

Figure S3. The gate leakage current $I_G$ and the drain current $I_D$ of devices without GNRs in the channel. (a) $I_D$ and $I_G$ versus $V_{GS}$ for a device with $L = 8$ nm. (b) $I_D$ versus $V_{DS}$ for a device with $L = 8$ nm. (c) $I_D$ and $I_G$ versus $V_{GS}$ for a device with $L = 15$ nm. (d) $I_D$ versus $V_{DS}$ for a device with $L = 15$ nm.

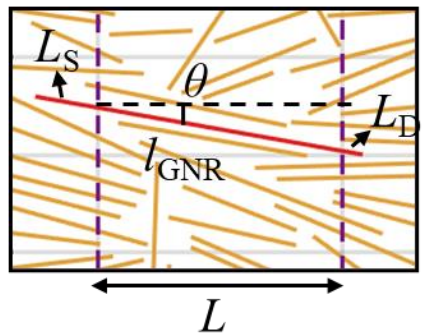

Figure S4. A representative simulated spatial distribution of 9-AGNRs. The purple dashed lines indicate the boundaries of the active device region (source and drain contacts). The solid lines are GNRs, and the line in red is a GNR that connects both the source and the drain contact. $L$ denotes the channel length; $l_{GNR}$ denotes the length of the GNR; $L_S$ and $L_D$ are the contact length on the source and the drain side, respectively; and $\theta$ is the angle of the GNR with respect to the direction of the channel.

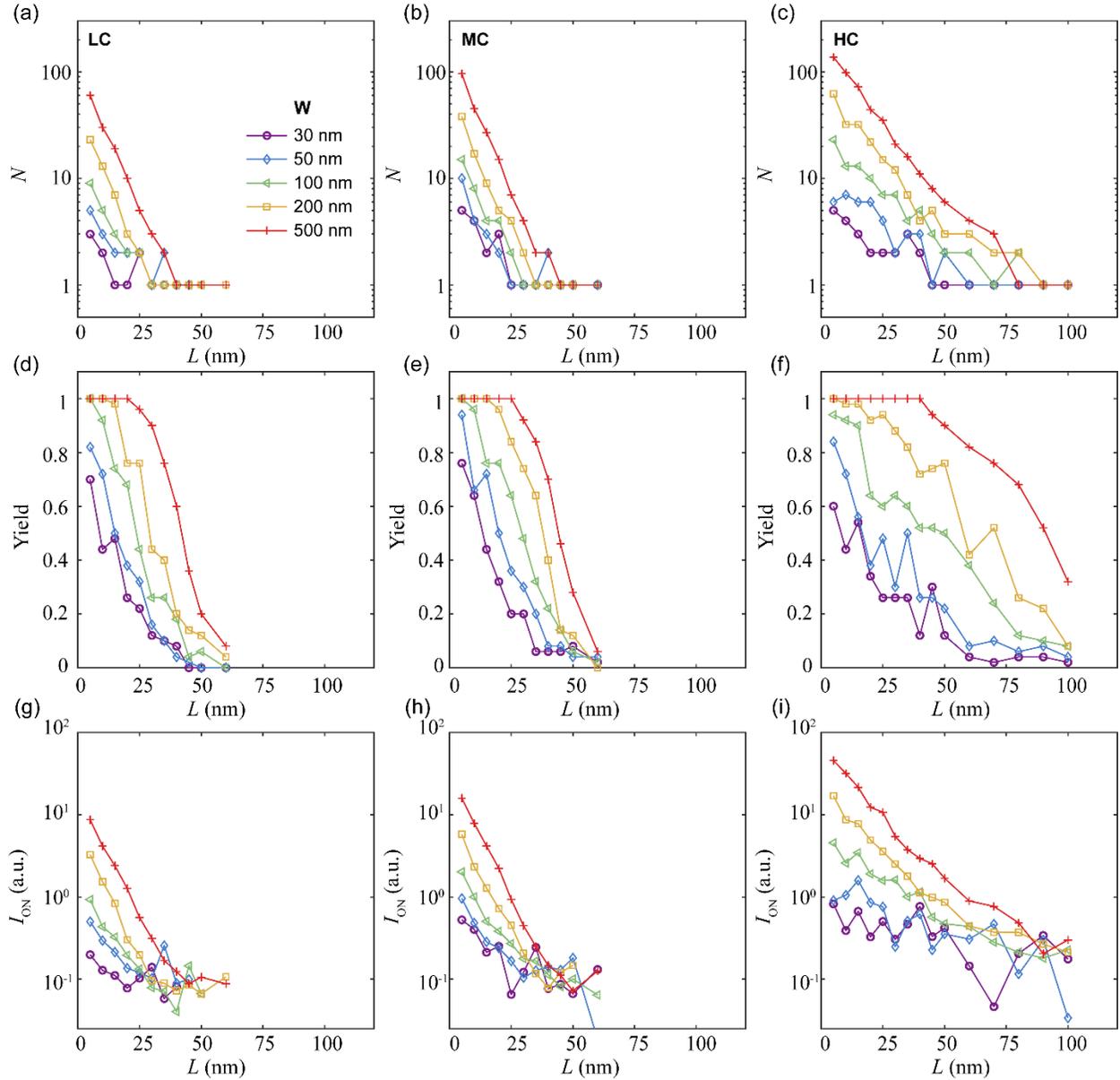

Figure S5. Additional simulation results of the number of connected GNRs ($N$), the device yield, as well as the on-state current ($I_{ON}$) for LC, MC, and HC samples with various channel lengths $L$ and channel

widths $W$. (a-c) $N$ versus $L$ with different $W$ for LC (a), MC (b), and HC (c) samples. (d-f) Yield versus $L$ with different $W$ for LC (d), MC (e), and HC (f) samples. (g-i) $I_{ON}$ versus $L$ with different $W$ for MC (g), MC (h), and HC (i) samples.

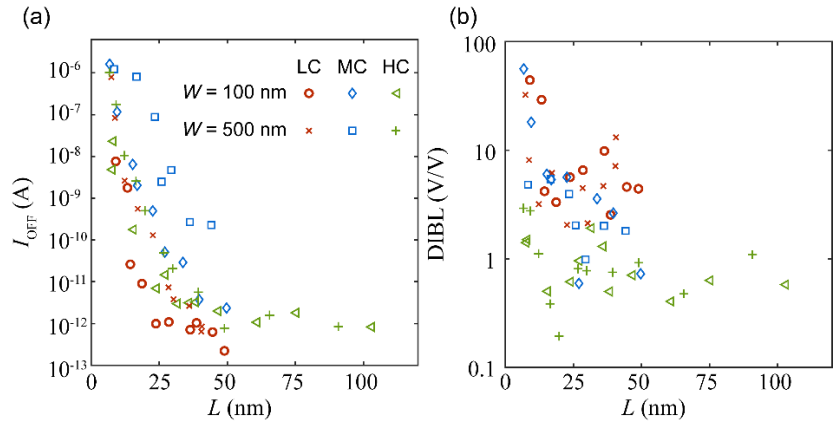

Figure S6. (a) Off-state current ($I_{OFF}$) and (b) drain induced barrier lowering (DIBL) as a function of the channel length $L$ of FETs with various channel widths $W$ and 9-AGNRs with different coverages.